\documentclass[%
reprint,
nofootinbib,
amsmath,amssymb,preprintnumbers,
aps,prl,twocolumn
]{revtex4-2}
\usepackage[T1]{fontenc}
\usepackage{enumitem}
 \usepackage{lmodern}
\usepackage{graphicx}
\usepackage{hyperref}
\usepackage{amsfonts}
\usepackage[named]{xcolor}

\usepackage{tocvsec2}
 \usepackage{slashed}
\usepackage{comment}

\newcommand\beq{\begin{eqnarray}}
\newcommand\eeq{\end{eqnarray}}

\newcommand{\half}{\frac{1}{2}}

\newcommand\eqn[1]{\label{eq:#1}} 
\newcommand\eq[1]{eq.~(\ref{eq:#1})} 
\newcommand\Eq[1]{Eq.~(\ref{eq:#1})}

\newcommand{\bfq}{{\mathbf q}}

\newcommand{\CA}{{\cal A}}

\newcommand{\CD}{{\cal D}}

\newcommand{\CI}{{\cal I}}
\newcommand{\CJ}{{\cal J}}

\newcommand{\CL}{{\cal L}}

\newcommand{\Tr}{{\rm Tr\,}}

\newcommand\vev[1]{\langle #1 \rangle}

\newcommand{\mybar}[1]%
        {\kern 0.6pt\overline{\kern -0.6pt#1\kern -0.6pt}\kern 0.6pt}

\def\Tr{\text{Tr}\,}

\def\half{\tfrac{1}{2}}

\hypersetup{
    colorlinks=true,   % false: boxed links; true: colored links
    linkcolor=cyan,     % color of internal links (box color = linkbordercolor)
    citecolor=blue,    % color of links to bibliography
    filecolor=magenta, % color of file links
    urlcolor=blue      % color of external links
}
\usepackage[section]{placeins}

\begin{document}

\preprint{INT-PUB-21-036}

\title{Index theorems, generalized Hall currents, and topology for gapless defect fermions}
\author{David B. Kaplan}
\email{dbkaplan@uw.edu}
\affiliation{Institute for Nuclear Theory, University of Washington, Box 351550, Seattle WA 98195-1550}
\author{Srimoyee Sen}
\email{srimoyee08@gmail.com}
\affiliation{Department of Physics and Astronomy,  Iowa State University, Ames IA 50011}

\begin{abstract}

We show how the index of the fermion operator from the Euclidean action can be used to uncover the existence of gapless modes living on defects (such as edges and vortices) in topological insulators and superconductors.  The 1-loop Feynman diagram that computes the index reveals an analog of the Quantum Hall current flowing on and off the defect -- even in systems without conserved currents or chiral anomalies -- and makes explicit the interplay between topology in momentum and coordinate space. We provide several explicit examples.
\end{abstract}

\maketitle

\section{Introduction}
\label{sec1}
Defects such as domain walls, vortices, and monopoles often host gapless boundstates when coupled to fermions  \cite{Jackiw:1975fn}, the existence of which can be related to topology and anomalies \cite{atiyah1968index,fujikawa1979path}.  However, their properties vary according to the dimension and symmetries of the bulk and defect theories.  In some cases, the existence of gapless modes can be deduced from currents flowing onto or off of the defect, as with the Integer Quantum Hall Effect or lattice domain wall fermions \cite{Kaplan:1992bt,jansen1992chiral}, while in other examples there are no such currents. Furthermore, the topology that governs the gapless states is in momentum space, and so the number and nature of such states can be sensitive to how the theory is regulated at short distance \cite{Golterman:1992ub, thouless1982quantized, niu1985quantized}. That too can depend on  the dimension of the bulk theory.

 In this paper we demonstrate a generic framework  which can be used to identify gapless defect modes in topological insulators and superconductors,  associating all of them with a generalized  Hall current with nonzero divergence flowing onto the defect, irrespective of whether the original theory contains any continuous internal symmetry, conserved currents, or chiral anomalies.  This approach involves computing the index  \cite{Callias:1977kg}  for the Dirac operator in the Euclidean action, where one has added  diagnostic background fields with nontrivial topology\footnote{The Callias index theorem has been discussed before in a different context, for determining time-independent solutions to the Dirac equation, see Refs.~ \cite{Weinberg:1981eu,Seiberg:2016rsg}. The Euclidean path integral has been used to investigate the role of global anomalies in topological materials \cite{witten2019anomaly}.}.    The index is determined by computing a one-loop Feynman diagram, where the integration over loop momentum can be directly related to topological properties of the fermion dispersion relation. We describe the method here and briefly give three explicit examples involving Dirac and Majorana fermions in two, three, and four spacetime dimensions.  A more detailed analysis, which includes consideration of the role of interactions, may be found in Ref.~\cite{longpaper}.

     The connection between massless states in Minkowski spacetime and the index of the Euclidean Dirac operator is not direct.  Consider the example of a Dirac fermion in two spacetime dimensions with  $\CD = \slashed{\partial} + m \epsilon(x)$, where a domain wall confines gapless states to the one-dimensional line $x=0$. In Minkowski spacetime this would correspond to a massless mode confined to the end of wire.  In Euclidean spacetime,  the only state annihilated by $\CD$ will be  one localized in the $x$ direction  but constant in Euclidean time $\tau$; as this is not a normalizable state it will not contribute to the index of $\CD$.  However, we can now imagine adding a second domain wall defect as a function of $\tau$; if there exists a gapless mode in the first place, it will now be localized in two dimensions, is not a zeromode of $\CD^\dagger$, and can contribute to the index of $\CD$, no matter how weakly the fermion interacts with that second domain wall. If we remove the original domain wall at $x=0$ eliminating the massless edge state of interest,  then the index vanishes. In this sense, the index of the modified theory reveals the gapless state in the original one.       More generally, one can reveal edge states by considering fermions propagating in arbitrary background fields. We will show that when our heuristic example of crossed domain walls is replaced by an arbitrary, smoothly varying complex scalar field one finds that the index is proportional to the field's vorticity. In higher dimensions localizing the zeromodes requires additional fields, such as gauge fields.  
     
Our approach then is to add extra diagnostic fields to the theory of interest and compute  the index of Euclidean spacetime operator $\CD$   in a derivative expansion, along the lines of Ref.~\cite{Goldstone:1981kk}.
We find that the index is proportional to a topological invariant of these fields in coordinate space, times a topological invariant constructed from the fermion dispersion relation in momentum space.  A nonzero value for the product of these winding numbers is taken to indicate the existence of massless states in the Minkowski version of the original theory.

As discussed in Refs.~\cite{Callias:1977kg,Weinberg:1981eu}, the index of a non-Hermitian elliptic operator $\CD$  can be defined as $\CI(0)\equiv \lim_{M\to 0}\ \CI(M)$, where
\beq
\CI(M) &=&\Tr \left(\frac{M^2}{\CD^\dagger \CD + M^2} - \frac{M^2}{\CD \CD^\dagger+M^2}\right)\cr
&=&\Tr \Gamma_\chi \frac{M}{K+M}\ ,
\eqn{IM}\eeq
where
\beq
K = \begin{pmatrix} 0 & -\CD^\dagger\\ \CD & 0\end{pmatrix},\ 
\Gamma_\chi  = \begin{pmatrix} 1 & 0\\ 0 & -1\end{pmatrix},\  \{K,\Gamma_{\chi}\}=0\ .
\eqn{Kdef}
\eeq
Let us now imagine that $S=\int \bar\psi \CD \psi$ is the Euclidean action for a system of interest with massless edge states in $(d+1)$-dimensional Minkowski spacetime, where we use the term ``edge states'' to refer massless fermions bound to a defect of any codimension.  Then $1/(K+M)$   looks like the propagator in a new theory with Euclidean action 
\beq
S =\int d^{d+1}x\, \mybar \Psi (K +M)\Psi\ ,
\eqn{act}
\eeq
where $\Psi$ is a complex fermion with its own continuous fermion number symmetry and, in the $M\rightarrow 0$ limit, a continuous axial symmetry, both of which are unrelated to any symmetries of the original theory. $\Psi$ also has twice as many components as the physical fermions. We can now express  $\CI(M)$ in terms of the new theory as
\beq
\CI(M) =-M\int d^{d+1}x \,\vev{\mybar\Psi(x)\Gamma_\chi  \Psi(x)}\ , 
\eqn{PS}
\eeq
where the quantum average is computed from a path integral over $\Psi$ and $\mybar\Psi$ with weight $e^{-S}$.  We are not restricted to an action like $S=\int \bar\psi \CD \psi$ which preserves fermion number; this analysis is relevant also for a Minkowski theory of real fermions with Euclidean action $\int \psi^T C \CD \psi$, where $C$ is the charge conjugation operator.

In the cases we will examine, $K$ will be a linear differential operator of the form $K = \Gamma_\mu \partial_\mu + V$, where the $V$ is some spacetime dependent matrix.   Then we can define the Noether current for the axial symmetry of \eq{act}, $\CJ_\mu = \mybar \Psi \Gamma_\mu \Gamma_\chi  \Psi$ which obeys an anomalous  Ward-Takahashi  identity
\beq
\partial_\mu \CJ_\mu = 2 M \mybar\Psi \Gamma_\chi  \Psi - \CA\  ,
\eqn{divJ}
\eeq
where the first term on the right is due to the explicit chiral symmetry breaking by $M$, and $\CA$ is the potential anomalous contribution due to the variance of the path integral measure \cite{fujikawa1979path}, with
\beq
\int d^{d+1}x\, \CA =-2 \lim_{\Lambda\to \infty}\Tr\Gamma_\chi  e^{K^2/\Lambda^2} =-2\, \CI(\infty)\ .
\eqn{fuji}
\eeq
It follows then from \Eq{PS} that
\beq
 \CI(M)= \CI(\infty) -\half \int d^{d+1}x\, \partial_\mu \vev{\mybar \Psi \Gamma_\mu \Gamma_\chi  \Psi}   \ .
\eqn{ind}
\eeq
Therefore, to compute the index $\CI(0)$ one need only compute the two terms on the right in the massless limit.  In all the cases we will consider,  the anomaly $\CA$ and hence $\CI(\infty)$ trivially vanish, and one need only compute the axial current flowing in from infinity, which requires computing  the one loop diagram for the chiral current  $ \vev{\mybar \Psi \Gamma_\mu \Gamma_\chi  \Psi}$ from the action in \Eq{act}. Note that in every case, a nontrivial index is associated with current inflow, independently from whether $\CD$ has a continuous symmetry or anomalies. This chiral current exists for every Minkowski theory and is unrelated to any chiral symmetry the original Minkowski theory might  have possessed; it behaves like a generalization of the familiar quantum Hall current and we will refer to it as such.   
%%%%%%%%%%%%%%%%%%%%%%%%%%%%%%%%%

\section{Majorana fermion in $1+1$ dimensions}

Our first example is a massive Majorana fermion in $1+1$ dimensions.  Our starting point is the Lagrangian in Minkowski spacetime,
\beq
\CL_M = \half\psi^T C \left(i\slashed{\partial}- m \right)\psi
\eeq
where $\psi$ is a real, two-component Grassmann spinor and we can take $\gamma^0= C = \sigma_2$, $\gamma^1 = -i\sigma_1$, $\gamma_\chi = \sigma_3$, where $\sigma_i$ are the Pauli matrices.
 This is the model considered in Ref.~\cite{Fidkowski:2009dba}, although here we do not include interactions, and we consider the case of infinite dimensions with a domain wall mass $m=m_0\epsilon(x^1)$ rather than a finite wire with two ends.  This theory has no continuous symmetries except Lorentz symmetry; it possesses the discrete spacetime symmetries $C$, $P$ and $T$ which play the roles as particle-hole, sublattice, and time reversal symmetries respectively in condensed matter systems. As is easy to see, a gapless fermion exists at $x^1=0$.

\begin{figure*}[t]
\centerline{\includegraphics[width=14 cm]{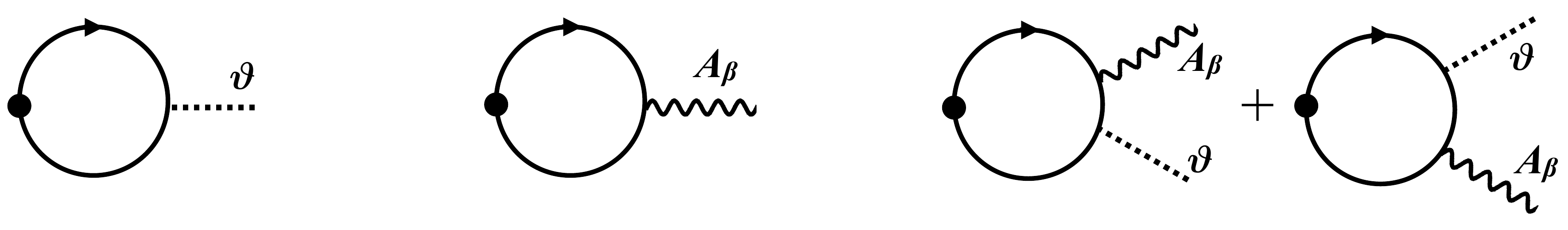}}
\caption{\it Loop diagrams for computing the generalized Hall current for the $1+1$ (left), $2+1$(center) and  $3+1$ (right two) dimension examples. The black dot is an insertion of the chiral current $\Gamma_\mu\Gamma_\chi$, and the propagators are $K^{-1}$.}
\label{fig:1}
\end{figure*}

The Euclidean Lagrangian is obtained from the Minkowski Lagrangian in $d+1$ dimensions as $\CL_E = -\CL_M$ along with the replacement $\partial_0\to i\partial_{0}$ and a redefinition of the $\gamma$ matrices so that they obey the $SO(d+1)$ Clifford algebra (we use a mostly minus metric).  In the present example with $d=1$ we have $\CL_E =\half  \psi^T C \CD\psi$, where $\CD = \slashed{\partial} + m$ with   $C = \gamma_0 = \sigma_2$, $\gamma_1 = -\sigma_1$,  and  $\gamma_\chi=\sigma_3$.  
Following the discussion in the introduction we generalize the model by replacing the mass by a scalar field $\phi_1$ and add a pseudoscalar field $\phi_2$ so that our Euclidean Dirac operator becomes
\beq
 \CD= \left(\slashed{\partial}+\phi_1+i \phi_2  {\gamma_\chi}\right)\ .
 \eqn{ddag}
\eeq

In order to compute the index of $\CD$  we next construct the Euclidean theory $\CL = \mybar\Psi (K+M)\Psi$  where $K$, is  specified by \Eq{Kdef}. The operator $K$ can be written as
\beq
K &=&   \sum_{\mu=0}^1 \Gamma_\mu \partial_\mu + i\phi_2\Gamma_2 + i \phi_1\Gamma_3
\eeq
where 
%we write $\phi_{1}+i\phi_2 = \rho e^{i\theta}$, and  our basis is
\begin{equation}
\begin{aligned}
%\beq
\Gamma_i &= \sigma_1\otimes \gamma_i\ ,\  
&\Gamma_2 &= \sigma_1\otimes \gamma_\chi\ , \cr 
\Gamma_3 &= -\sigma_2\otimes 1\ , \ 
&\Gamma_\chi &= \sigma_3\otimes 1\ ,
%\ ,\cr
%\Sigma_{34} &= \frac{i}{4}\left[\Gamma_3,\Gamma_4\right] = \half\sigma_3\otimes\sigma_3\ .
%\eeq
\end{aligned}
\end{equation}
with $i=0,1$.  
The leading contribution to the current $\CJ_\mu$ in a derivative expansion of the scalar fields is shown in Fig.~\ref{fig:1} and is  proportional to $\partial\theta$, where we write $\phi = \phi_1 + i\phi_2 = v e^{i\theta}$ and expand about $\theta=0$. To linear order the $\theta$ vertex is given by $-iv \Gamma_2$. We define $K_0=K(\theta=0)$. Taking $M\to 0$,  the Feynman diagram yields
 \beq
\CJ_\mu&=&v\, \frac{\partial\theta}{\partial x_\nu}\int \frac{d^2q}{(2\pi)^2}\,\Tr\Biggl[
\Gamma_\mu\Gamma_\chi 
 \left(\frac{\partial  \tilde K_0^{-1}}{\partial q_\nu} \right)\Gamma_2\tilde K_0^{-1} \Biggl]\cr
 &=&
 \epsilon_{\mu\nu} \partial_\nu \theta \int \frac{d^2q}{(2\pi)^2}\frac{4v^2}{(q^2+v^2)^2} \cr 
 &=&
  \frac{1}{\pi} \epsilon_{\mu\nu} \partial_\nu \theta  \ .
\eqn{index2}
\eeq
The index is then computed to be
\beq
\text{ind}(\CD)=\CI(0) = -\half\int d^2x \,\partial_\mu\CJ_\mu = \oint \frac{d\theta}{2\pi} = -\nu_\phi\ ,
\eqn{res2}\eeq
where $\nu_\phi$ is the winding number of the scalar field $\phi$ in the $x_0-x_1$ plane.  This result is consistent with the heuristic example discussed in the introduction of the crossed domain wall configuration $\phi_1 = m\epsilon(x_1)$, $\phi_2=\mu\epsilon(x_0)$ with $m>0$, $\mu>0$, for which one finds $\nu_\phi=-1$. We see that the index of the Euclidean Dirac operator $\CD$ reveals the existence of massless edge state in the presence of nontrivial spatial topology for the background $\phi$ field.  

The above calculation does not fully reveal the topological nature of the edge state, however,  and one might still ask as to why the Feynman integral over momentum results in an integer for the index rather than some arbitrary real number.  
To address this we borrow the techniques of Ref.~\cite{Golterman:1992ub, ishikawa1987microscopic} to show that the Feynman integral is actually computing a winding number. The result in \eq{index2} is unchanged if we substitute for the skew diagonal blocks in $K_0$ 
\beq
\CD_0\rightarrow \xi \equiv  \frac{\CD_0}{\sqrt{\det\CD_0}} =   \frac{v }{\sqrt{q^2+v^2}} +i \hat q_\mu\gamma_\mu\frac{q}{\sqrt{q^2+v^2}} \ ,
\eqn{xi}
\eeq
where $\CD_0 = \CD\vert_{\theta=0}$, while $\CD_0^\dagger$ is replaced by $\xi^{\dagger}$. The matrix $\xi$ is   unitary and the generalized Hall current in \eq{index2} may be written in terms of $\xi$ as
\beq
\CJ_\mu&=&\frac{i}{2}\epsilon_{\mu\nu}\partial_\nu \theta\,\epsilon_{\sigma\tau}
\cr
&&\times\int \frac{d^2q}{(2\pi)^2}\Tr\gamma_\chi\biggl[\left(\xi^\dagger \partial_\sigma \xi\right)\left( \xi^\dagger\partial_\tau  \xi \right)
 +\left(\xi \partial_\sigma \xi^\dagger\right)\left(\xi \partial_\tau  \xi^\dagger\right)   \bigr]\ .\cr&&
\eqn{D3new}
\eeq
The momentum integral can be shown to be the winding number associated with the map $U(q) = \xi^2(q)$ from momentum space to $S^2$ (see Ref.~\cite{longpaper} for details), where
\beq
U = \frac{v^2-q^2}{v^2+q^2} + \frac{2 i v \slashed{q}}{v^2+q^2}\equiv a(q) + i \slashed{b}(q)\ .
\eeq
Since   $a^2 + b_i b_i = 1$, $U$ describes  a unit 3-vector, which lives on $S^2$.  We see that all possible points on $S^2$ correspond to a unique value of the 2-momentum $q_i$, except that the limit of infinite momentum in any direction gets mapped to $U=-1$.  So momentum space has itself been compactified to $S^2$, and $U$ describes a nontrivial map in the homotopy group $\pi_2(S^2) = {\mathbb Z}$.  The winding number $\nu_q=1$ of this map is computed by the above integral, and we have the result for the index
\beq
\text{ind}(\CD)=-\nu_\phi\nu_q\ .
\eeq
Now the full topological meaning of the index is manifest and it is evident how a Feynman diagram can produce an integer.  In general the index can change only at values for the  parameters in the theory when our substitution in \eq{xi} fails,  namely where $\det\CD_0$ vanishes for some momentum; such singularities are equivalent to the bulk gap vanishing, allowing the zeromode to delocalize, and can only happen in this case when $v=0$.

As a last comment on this model, the anomaly term $\CA$ in \eq{fuji} vanishes because the trace over momenta gives a factor of $\Lambda^2$ in two dimensions, while a nonzero $\Gamma$ matrix trace requires two powers of $K^2/\Lambda^2$, since the $\Gamma$ matrices behave like those for $SO(4)$, and hence $\CA$ vanishes in the limit $\Lambda\to \infty$. 
%
%%%%%%%%%%%%%%%%%%%%%%%%%%%%%%%%%%

 \section{Topological insulator in $3+1$ dimensions}
 
 %%%%%%%%%%%%%%%%%%%%%%%%%%%%%%%%%%

We next compute the index for a topological insulator in $3+1$ dimensions, consisting of a Dirac fermion with a domain wall mass \cite{Seiberg:2016rsg}.  The domain wall in this case is a $2+1$ dimensional defect hosting massless fermions. In order to construct the current-inflow picture we consider the following Euclidean Lagrangian
\beq
\CL_E =\bar\psi \CD \psi\ ,\qquad \CD =  \slashed{D}+ \phi_1 + i \phi_2\gamma_5  \ ,
\eeq
 where $\phi_1$ takes the place of the Dirac mass, and $\phi_2$ is a diagnostic background scalar field needed to localize the edge mode as in the previous example.
More background fields are required in higher dimensions to localize a zeromode, and to that end we include an Abelian gauge field in addition to $\phi_{1,2}$.    The operator $K$ in \eq{Kdef} is given by
\beq
K &=& \begin{pmatrix} 0 & -\CD^\dagger\\ \CD & 0 \end{pmatrix} = \sum_{a=0}^3 \Gamma_a D_a +i \phi_2 \Gamma_4 +i \phi_1  \Gamma_5\ ,
%\cr
%&=&
%e^{ -i \theta \Sigma_{56}}K_0 e^{i \theta \Sigma_{56}},
%\cr
%K_0&=& \Gamma_a D_a -i \partial_\mu\theta\Gamma_\mu\Sigma_{56}-i \rho  \Gamma_6\ ,
\eeq
our basis for the $8\times 8$ $\Gamma$-matrices being
\begin{equation}
\begin{aligned}
\Gamma_a &= \sigma_1\otimes \gamma_a\ ,\quad &a=0,\ldots,3\ ,\\
\Gamma_4 &= \sigma_1\otimes \gamma_5 & \Gamma_5  = \sigma_2\otimes 1\ , 
%\\
%\Sigma_{56}&= \frac{i}{4}[\Gamma_5,\Gamma_6] = -\half\sigma_3\otimes\gamma_5\ .
\end{aligned}
\end{equation} 
 with $\Gamma_\chi =  \sigma_3\otimes 1$.   We must now compute the divergence of the chiral current, $\vev{\mybar\Psi\Gamma_\mu\Gamma_\chi\Psi}$, where again the mismatch between the four spacetime dimensions, and $8\times 8$ $\Gamma$-matrices ensures that the anomaly $\CA$ in \eq{fuji} vanishes. Computing in a derivative expansion requires evaluating the two diagrams to the right in \ref{fig:1}, with the $M=0$ result
 \beq
 \CJ_\mu=\vev{\mybar\Psi\Gamma_\mu\Gamma_\chi\Psi} = \frac{\nu_q}{2\pi^2} \epsilon_{\mu\alpha\beta\gamma} F_{\alpha\beta}\ \partial_\gamma\theta\ ,
\eqn{4dres} \eeq
 with $\phi = \phi_1 + i \phi_2=\rho e^{i\theta}$.  Similar to the previous example, $\nu_q=1$ is a winding number computed by the Feynman loop integral of the map from momentum space to to the space of spinor orientations defined by the bulk fermion propagator.  In this case we find the map to be an element of $\pi_4(S^4)={\mathbb Z}$ \cite{longpaper}.
 
 When we consider the case of the domain wall mass profile in the original theory, we have $\phi_1 = m_0 \epsilon(x_3)$; a suitable background field configuration for $\phi_2$ and ${\bf A}$ to localize a massless  edge state is the monopole configuration discussed in Ref.~\cite{cheng2013fermion} with couplings set to $e=1$ and $g=2\pi$:
  \beq
 {\bf A }= -\frac{(1+\cos\theta){\bf e}_\varphi}{2 r\sin\theta}\ ,\qquad \phi_2 =\alpha -\frac{1}{2r}\ ,
 \eqn{CF}
 \eeq
 where $\{r,\theta,\varphi\}$ are polar coordinates for the Euclidean space spanned by $\{x_0,x_1,x_2\}$.  In this background we can compute the index $-\half\int d^4x\,\partial_\mu \CJ_\mu $, finding
 \beq
 \int d^4x\,\partial_\mu \CJ_\mu = 
 %2\,\theta(\mu)
 \left(1+\frac{\alpha}{|\alpha|}\right)\ \Longrightarrow\  \text{ind}(\CD) =- \theta(\alpha)\ .
 \eeq
 This nontrivial index indicates the existence of gapless edge states: a transition in a topological quantity such as the gap is only possible when fields become delocalized, and so we see that happens at $\alpha=0$, indicating no other scale exists in the low energy spectrum of the theory.   Our result \eq{4dres} can equally be applied to identify a massless fermion bound to a magnetic monopole, by adding an external scalar field in the form of a vortex.  The index is negative when nonzero because the field configuration in \eq{CF} causes $\CD^\dagger$ to have a zeromode when a massless edge state is present, instead of $\CD$.
  
 %%%%%%%%%%%%%%%%%%

\section{Majorana fermions in $2+1$ dimensions}

For our third example we next consider a two component complex fermion $\psi$ in $2+1$ dimensions with both a constant Majorana mass term $\mu$ and a real Dirac mass term  $m(x^1)$. This theory is known to describe chiral topological superconductors \cite{PhysRevB.82.184516, Lian10938, PhysRevB.61.10267}. Nonzero $\mu$ breaks the  $U(1)$ fermion number symmetry of the Dirac theory to  $Z_2$. When $m(x^1)$ has a domain wall profile there appear zero, one, or two massless   Majorana-Weyl fermions on the defect, depending on the ratios $m_{\pm}/\mu$, where $\pm m_\pm$ are the asymptotic values of $m(x^1)$ at $x^1=\pm \infty$.

In Minkowski spacetime the Lagrangian for this system may be written as 
\beq
\CL_M = \bar\psi \left(i\slashed{\partial} - m\right)\psi + i\frac{\mu}{2}\psi^T C \psi+i\frac{\mu}{2}\bar \psi \,C\bar \psi^T \ ,
\eqn{majmod}
\eeq
 where $M$ is real, $\mu$ is real and positive, and we can work in the explicit basis $\gamma^0 = \sigma_2$, $\gamma^1=-i\sigma_1$, $\gamma^2 = i\sigma_3$ and $C=\sigma_2$. After rotating to Euclidean spacetime, decomposing into its real and imaginary parts $\psi = \chi_1 + i \chi_2 $ with real 2-component spinors $\chi_i$, and then defining $\zeta_\pm = (\chi_1\pm\chi_2)/\sqrt{2}$, one can write the Euclidean Lagrangian as
 \beq
 \CL_E &=& \half\left[\zeta_+^T C \CD_+ \zeta_+ +\zeta_-^T C \CD_- \zeta_- \right]\ , \ \CD_\pm = \slashed{\partial} + (m\pm \mu)\ ,\cr &&
\eqn{D2}
 \eeq
where the gamma matrices are given by $C=\gamma_0=\sigma_2$, $\gamma_1 = -\sigma_1$, $\gamma_2 = \sigma_3$.  The index will then be the sum of the indices of $\CD_+$ and $\CD_-$.  To compute them we add a gauge field as the diagnostic field, construct $K_\pm$ from $\CD_\pm$, and compute the generalized Hall current from the second Feynman diagram in Fig.~\ref{fig:1} to leading order in a derivative expansion.  Consider the case $\mu=0$; then the   current for either of the $\zeta_\pm$ may be written as \cite{longpaper}
\beq
 \CJ_\alpha 
%&=& - i p_\gamma \tilde A_\beta(p) \int \frac{d^3q}{(2\pi)^3} \, \Tr \left[\Gamma_\chi \left(\tilde K_0^{-1} \partial_\gamma \tilde K_0\right) \left(\tilde K_0^{-1} \partial_\beta \tilde K_0\right) \left(\tilde K_0^{-1} \partial_\alpha \tilde K_0\right)\right]\cr
&=&
 -\epsilon_{\alpha \beta\gamma}\partial_\gamma   A_\beta  \left(\frac{1}{3}\epsilon_{ijk}\right)\cr &&\times
\int \frac{d^3q}{(2\pi)^3}  \Tr \left[ \left(\tilde \CD_0^{-1} \partial_i\tilde\CD_0\right) \left(\tilde \CD_0^{-1} \partial_j \tilde \CD_0\right) \left(\tilde \CD_0^{-1} \partial_k \tilde \CD_0\right)\right] \cr
 &&\cr &&
 \eqn{Dloop}\eeq
where for $\CD_0$ we take $\CD_\pm\vert_{{A_\mu}=0}$.  To understand the underlying topology we can define 
\beq
 U(\bfq) = \frac{\tilde \CD_0(\bfq)}{\sqrt{\det  \tilde\CD_0(\bfq)}}  \equiv \cos\frac{\theta}{2}+ i \hat{ \slashed{\theta}}\cdot{\boldsymbol\gamma}\sin\frac{\theta}{2}
\eqn{udef1}
\eeq
where ${\boldsymbol \theta}$ is a real 3-vector with $\theta= |{\boldsymbol \theta}| $ and
\beq
 \cos\frac{\theta}{2}= \frac{m}{\sqrt{m^2+q^2}} \ ,\quad
\sin\frac{\theta}{2}=\frac{ q}{\sqrt{m^2+q^2}}\ ,\quad  \hat {\theta}=\hat{\bf q}\ . \cr&&
\eqn{udef2}\eeq
which allows us to rewrite the expression for the current as
\beq
 \CJ_\alpha &=&
-\frac{1 }{\pi} \epsilon_{\alpha \beta\gamma}\partial_\gamma  A_\beta \left(\frac{1}{24\pi^2}\epsilon_{ijk}\right)
\cr &&
\times \int_{\theta\le \pi } d^3\theta\, \Tr \left[ \left( U^\dagger \partial_i  U \right)
 \left(  U^\dagger   \partial_j  U\right) 
 \left( U^\dagger \partial_k  U\right)\right] \ , \cr&&
\eqn{U3}
 \eeq
with $\partial_iU\equiv  \partial  U/\partial \theta_i$.  The integral looks like the winding number of a map from momentum space compactified to $S^3$, to $SU(2)\cong S^3$, except that the integral is only over half of $S^3$.  The problem can be seen in \eq{udef1}: $\cos\theta/2 \ge 0$ for all momenta, so $U$ cannot take on all values in $SU(2)$.   The problem is solved when the theory is regulated.  With a Pauli-Villars regulator one substitutes $\CD(m) \to \CD(m)/\CD(\Lambda)$ and takes the $\Lambda\to\infty$ limit.  The result is that $U\to U_\text{reg}$  in \eq{U3} where
\beq
 U_\text{reg}(\bfq) &=& \frac{ \tilde  \CD_\text{reg}(\bfq)}{\sqrt{\det  \tilde  \CD_\text{reg}(\bfq) }}    
 \equiv \cos\frac{\theta_\text{reg}}{2}+ i \hat {\boldsymbol \theta}_\text{reg}\cdot{\boldsymbol\gamma}\sin\frac{\theta_\text{reg}}{2} \cr&&
   \eeq
   where $ \hat {\boldsymbol \theta}_\text{reg}=\hat{\bf q}$ as before, and
   \beq
   \cos\frac{\theta_\text{reg}}{2}= \frac{\Lambda  m+q^2}{\sqrt{\left(m^2+q^2\right)
   \left(\Lambda ^2+q^2\right)}}\ .
     \eeq
Now we have $U_\text{reg}(\bfq)\xrightarrow{q\to\infty} = 1$,  while $U_\text{reg}(0) = \text{sgn}(m\Lambda)$.  This regulated theory describes a well defined map $S^3\to S^3$ which is nontrivial if $m$ and $\Lambda$ have the opposite signs, and trivial  if they don't.  On replacing $m\to m(x) \pm\mu$ and integrating the divergence of the generalized Hall current over Euclidean spacetime, we arrive at the result for the index for the whole system
\beq
\text{ind}(\CD) = \nu_A \nu_q\ ,
\eeq
where 
\beq
\nu_A &=&   \frac{1}{2\pi} \oint {\bf A}\cdot d{\bf \ell}\ ,\cr
 \nu_q &=& \theta(m_+ + |\mu|) + \theta(m_+ - |\mu|) \cr &&-\theta(m_- + |\mu|) - \theta(m_- - |\mu|)\ .
 \eeq
 where we have assumed that $\mu$ is spatially constant while $m(x)\xrightarrow{x\to\pm\infty}  m_\pm$. Note that $\nu_q$ can take on the values $0,\pm1,\pm 2$ depending on the relative values of $m_\pm$ and $\mu$.  Once can verify that this result agrees with explicit edge state solutions to the equations of motion \cite{longpaper}, and we see that the generalized Hall current is sensitive to topological phase transitions as one varies parameters in the theory.

%%%%%%%%%%%%%%%%

 \section{Discussion}
 
We have shown how gapless fermions modes bound to defects or solitons in various dimensions may be detected by computing the index of the Euclidean Dirac operator in the presence of additional background fields.  The  method  involves determining the divergence of a generalized Hall current via a 1-loop Feynman integral, which calculates a topological winding number of the fermion propagator, the field theoretic generalization \cite{Jansen:1992tw,Golterman:1992ub}  of the TKNN result \cite{thouless1982quantized}.  Regularization is required to make topological sense of the result in odd spacetime dimensions. These currents can be computed for systems without chiral symmetries or anomalies, and generalize the concept of the Hall current. The examples considered here are in the $BDI$, $D$ and $DIII$ topological  classes in one, two and three spatial dimensions respectively; each is known to have a topological invariant taking values in ${\mathbb Z}$, and so perhaps it is not surprising that in each case we find momentum space topology governed by the homotopy groups $\pi_n(S^n)={\mathbb Z}$.  However, in Ref.~\cite{longpaper} we show this method correctly identifies the edge state spectrum for the $D$ class in one spatial dimension, with topological invariant ${\mathbb Z}_2$, yet surprisingly, even in that case the momentum space topology of the generalized Hall current is given by $\pi_n(S^n)$. It remains to be seen how comprehensive our approach is, whether it can be applied to theories with interactions, and whether this generalized Hall current has any experimental implications in Minkowski spacetime.\\ 

%%%%%%%%%%%%%%%%%%%%%

\section{acknowledgements}
We thank Biao Lian for introducing us to the literature on chiral topological superconductors. 
DBK is supported in part by DOE Grant No. DE-FG02-00ER41132 and  by the DOE QuantISED program through the
theory  consortium ``Intersections of QIS and Theoretical Particle
Physics'' at Fermilab. SS is supported by U.S. Department of Energy, Office of Science, Office of Nuclear Physics Quantum Horizons program under the Early Career Award DE-SC0021892.

%\bibliographystyle{utphys}
%\bibliography{anisotropic}
\bibliography{index2}
\bibliographystyle{JHEP}

 \end{document}